\documentclass[aps,prl,twocolumn,superscriptaddress,altaffillsymbol]{revtex4-1}

\usepackage{graphicx}
\usepackage{subfigure}
\usepackage{xcolor}
\usepackage{multirow}

\usepackage{dcolumn}

\UseRawInputEncoding

\newcommand{\fix}[1]{{\color{black} #1}}

\newcommand{\micron}{$\mu$m}

\usepackage{lineno}
\begin{document}

%Title of paper
%\title{Increasing Effective Laser Intensity Using Compound Parabolic Concentrator Targets}
\title{Order of magnitude increase in laser-target coupling at near-relativistic intensities using compound parabolic concentrators}
%incrased target coupling/electron temperature measurements, performance metrics 
%Observation of Order of magnitude increase in laser-to-target coupling via new acceleration mechanism at ARC
%Increasing Effective Laser Intensity Using Compound Parabolic Concentrator Targets
%}

% Primary Authors
\author{G.\ J.\ Williams }
\email{williams270@llnl.gov}
\affiliation{Lawrence Livermore National Laboratory, Livermore, California 94550, USA}

\author{A. Link}
\affiliation{Lawrence Livermore National Laboratory, Livermore, California 94550, USA}

\author{M. Sherlock}
\affiliation{Lawrence Livermore National Laboratory, Livermore, California 94550, USA}

% Alphabetical:
%
% A - E

\author{D.\ A.\ Alessi}
\affiliation{Lawrence Livermore National Laboratory, Livermore, California 94550, USA}

\author{M. Bowers}
\affiliation{Lawrence Livermore National Laboratory, Livermore, California 94550, USA}

% F - J

\author{B. P. Golick}
\affiliation{Lawrence Livermore National Laboratory, Livermore, California 94550, USA}

\author{M. Hamamoto}
\affiliation{Lawrence Livermore National Laboratory, Livermore, California 94550, USA}

\author{M. R. Hermann}
\affiliation{Lawrence Livermore National Laboratory, Livermore, California 94550, USA}

% K - O
\author{D. Kalantar}
\affiliation{Lawrence Livermore National Laboratory, Livermore, California 94550, USA}

\author{K. N. LaFortune}
\altaffiliation[Present address: ]{LCLS, SLAC National Accelerator Laboratory, Menlo Park, California 94025 USA}
\affiliation{Lawrence Livermore National Laboratory, Livermore, California 94550, USA}

\author{A. J. Mackinnon}
\affiliation{Lawrence Livermore National Laboratory, Livermore, California 94550, USA}

\author{A.\ MacPhee}
\affiliation{Lawrence Livermore National Laboratory, Livermore, California 94550, USA}

\author{M. J.-E. Manuel}
\affiliation{General Atomics, San Diego, California 92186, USA}

\author{D. Martinez}
\affiliation{Lawrence Livermore National Laboratory, Livermore, California 94550, USA}

\author{M. Mauldin}
\affiliation{General Atomics, San Diego, California 92186, USA}
%mauldin@fusion.gat.com

% P - T

\author{L. Pelz}
\affiliation{Lawrence Livermore National Laboratory, Livermore, California 94550, USA}

\author{M. Prantil}
\affiliation{Lawrence Livermore National Laboratory, Livermore, California 94550, USA}

\author{M. Quinn}
\affiliation{General Atomics, San Diego, California 92186, USA}

\author{B. Remington}
\affiliation{Lawrence Livermore National Laboratory, Livermore, California 94550, USA}

\author{R. Sigurdsson}
\affiliation{Lawrence Livermore National Laboratory, Livermore, California 94550, USA}

% U - Z 

\author{P. Wegner}
\affiliation{Lawrence Livermore National Laboratory, Livermore, California 94550, USA}

\author{K. Youngblood}
\altaffiliation[Present address: ]{Lawrence Livermore National Laboratory, Livermore, California 94550, USA}
\affiliation{General Atomics, San Diego, California 92186, USA}

% Senior Author

\author{Hui Chen}
\affiliation{Lawrence Livermore National Laboratory, Livermore, California 94550, USA}

\date{\today}
	
\begin{abstract}
%\fix{Nature-style abstract:} \\
Achieving a high conversion efficiency into relativistic electrons is central to short-pulse laser application and fundamentally relies on creating  interaction regions with intensities ${\gg}10^{18}$~W/cm$^2$.
Small focal length optics are typically employed to achieve this goal; however, this solution is impractical for large kJ-class systems that are 
constrained by facility geometry, debris concerns, and component costs.
We fielded target-mounted compound parabolic concentrators to overcome these limitations and achieved nearly an order of magnitude increase to the conversion efficiency and more than tripled electron temperature compared to flat targets.
Particle-in-cell simulations demonstrate that  plasma confinement within the cone and formation of turbulent laser fields that develop from cone wall reflections  are responsible for the improved laser-to-target coupling. 
{These passive target components can be used to improve the coupling efficiency for all high-intensity short-pulse laser applications, particularly at large facilities with long focal length optics.}

\end{abstract}

\maketitle

% % % % % %
% Introduction 
% % % % % % 

%  		What are we talking about and what is the application?
Intense, short pulse lasers  have demonstrated wide utility for ion acceleration \cite{Fuchs:2006kq}, isochroic heating of matter into exotic states \cite{PhysRevLett.91.125004}, and generating positron-electron antimatter pairs \cite{PhysRevLett.102.105001}.  
%		What are the requirements?
These applications require the generation of relativistic electrons with temperatures on the order of 1-10~MeV and typically necessitate a laser driver with ultra-relativistic intensities $I_L\lambda^2  \geq 10^{19}$~W/cm$^2$~$\mu$m$^2$, where $I_L$ is the laser intensity and $\lambda$ the  wavelength.  
% 		How does it work?
For solid targets at these intensities, electrons are accelerated through direct interaction with the laser field via the ponderomotive force \cite{wilks1992absorption} and those that slowly dephase or observe a significant number of laser oscillations reach the highest energies.
%
% you can tailor the acceleation mechanism here // optimize the source // the electron accleration can be optimzed.
The electron acceleration can be optimized by tailoring preplasma scale length \cite{Pukhov1999Scaling,Peebles_2017} using structured target interfaces \cite{Purvis:2013Relativistic,PhysRevLett.116.085002}, encouraging laser-channeling \cite{doi:10.1063/1.4946024}, using multi-picosecond laser pulse durations \cite{PhysRevLett.109.195005}, or by having large focal spots \cite{Williams2020ElectronsARC,Kemp2020Electrons}.  
%		What is the problem?
However, intrinsic to these or any other laser-target interactions is the low-intensity wings of a realistic laser spot that do not contribute to the acceleration of relativistic electrons and therefore reduces the coupling.  %the inefficient and
Recapturing this energy has the potential for significant enhancement of laser absorption.
%
 %is the low-intensity wings of the laser pulse that do not contribute to 
%
 %inefficiency any of these interactions
%The region outside the central hot spot of the laser is typically thought to be wasted energy and is a significant loss of efficiency in any short-pulse application.
%		

Focal profiles at mid- and large-scale laser systems are typically far from the diffraction limit due to phase distortions incurred during the main laser amplification. 
%Inefficient focal profiles  mid- and large-scale laser systems, the full width at half maximum spot size typically amounts to $\ll$50\% of the total energy in the beam due to phase distortions that increase the spot from the diffraction limit. 
%fraction of the laser capable of reaching relativistic intensities is typically $\leq$50\% due to phase distortions that increase the spot from the diffraction limit. 
%		
In addition, physical constraints of a large laser facility, such as target debris, hardware interference concerns, and limited chamber geometries necessitate large focal length final optics.
For example, at the Advanced Radiographic Capability (ARC) laser \cite{Nicola2015ARC}  within the National Ignition Facility (NIF) \cite{doi:10.13182/FST15-144},  focusing optics are external to the 10~m diameter target chamber, achieving a maximum intensity of ${\sim}$1-3 $\times 10^{18}$~W/cm$^2$. %		What is the solution?
As this is below the optimal regime for many short-pulse applications, enhancing the laser-target coupling is therefore critical to enabling such capabilities at the NIF.

%schemes to enhance the coupling to the target are required such as using a passive target-mounted optic to recapture the otherwise unused energy from the wings of the laser spot.

%The relatively large spot, long pulse duration, and energy contrast 

% drastically increase coupling efficiency and energy transfer to secondary particles.

% % % % % % % % % % % % % % 
% Overview of the work and results:
% % % % % % % % % % % % % % 

Here, we report on performance improvements in near-relativistic laser-matter interactions using compound parabolic concentrator (CPC) cone targets. Several experimental variables, including concentrating  geometries, laser pulse duration, and number of incident beams, were investigated to optimize the interaction and  resulted in an increase in the electron conversion efficiency of ${>}7\times$ and temperature of ${>}3\times$ compared to a flat target interaction. CPC cone targets also enabled the first observations of positron-electron pairs at the NIF. 
%
 %which resulted in an increase the electron conversion efficiency of ${>}7\times$ and temperature of ${>}3\times$ compared to a flat target interaction. 
%
%
%that resulted in the conversion efficiency into electrons and \fix{x-rays} was experimentally demonstrated to increase by up to 5$\times$ and electron temperature increased by ${>}2\times$. 
%
The measured pair yield exceeded expectations by 10$\times$ from established scaling with laser intensity, indicating a strong enhancement in the acceleration of ${>}10$~MeV electrons.
The electron temperatures were 10-20${\times}$ higher than predicted from ponderomotive scaling. 
These performance metrics are generally associated with laser intensities on the order of $10^{19}$-$10^{20}$~W/cm$^2$ \cite{chen2015scalingPoP}, more than an order of magnitude above experimental intensities.
CPC designs have been widely used in low-intensity applications to increase fluence \cite{WINSTON197489}. A ray trace of the CPC cones  designed for the ARC laser geometry suggest that the large fraction of energy contained in the wings of the beam (${\sim}80\%$) could be repointed towards the central beam spot \cite{MacPhee2020CPC}. 
%
%These results are consistent with lasers 
%Using ponderomotive arguments, this is equivalent to having a laser with intensities increasing the effective laser intensity of NIF-and co by ${\sim}10\times$ to ${\sim}$2 $\times 10^{19}$~W/cm$^2$. 
At high intensities, we show throught hydrodynamic and particle-in-cell (PIC) simulations that while the cone provides a modest enhancement in laser intensity compared to a flat target, the increased performance is primarily due to the presence of turbulent laser fields from light reflecting from the cone walls as well as increased plasma volume (scale length) that develops due to the focusing and confining properties of the cone. 
CPC cone targets have demonstrated coupling enhancement on the NIF-ARC laser and are potentially compatible with any system, particularly those with a long focal length or low-Strehl profile.
%\fix{while this has been demonstrated on NIF-ARC, these targets can likely be used to enchance the coupling efficiency in any long f-number system}. 

% % % % % % % % % % % % % % % %
% Experimental Setup / Cone Geometry:
% % % % % % % % % % % % % % % %

\begin{figure}[t]
\subfigure{\label{fig:ConePicture}\includegraphics[width=0\textwidth]{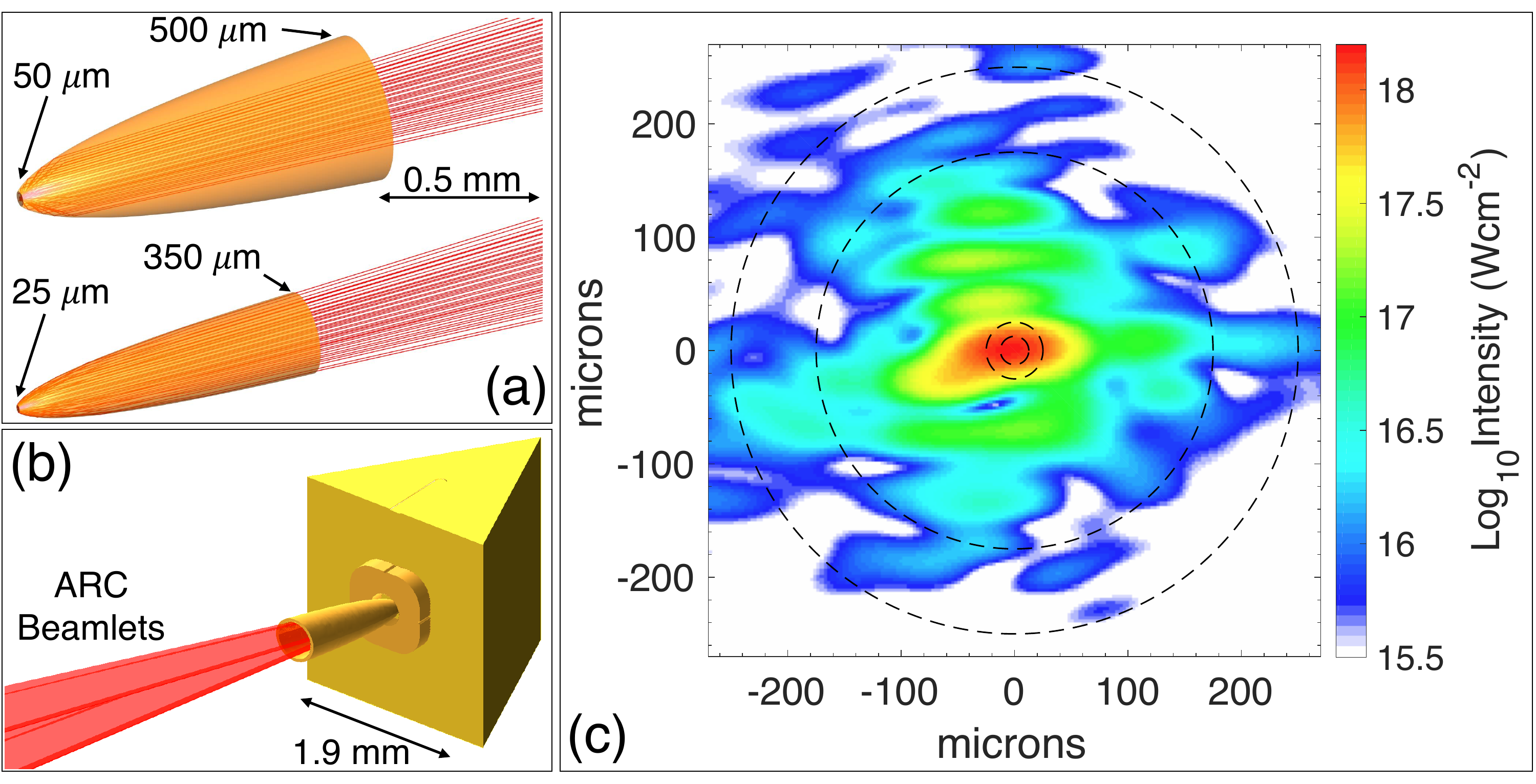}}   
\subfigure{\label{fig:TargetPicture}\includegraphics[width=0\textwidth]{Fig1_Cone_Spot_v3.pdf}}  
\subfigure{\label{fig:LaserSpot}\includegraphics[width=0.48\textwidth]{Fig1_Cone_Spot_v3.pdf}}  
\caption{(a) Diagram of CPC cones with 50~\micron{} diameter cone tip (top) and 25~\micron{} diameter cone tip (bottom). 
(b) Au prism target. 
(c) Modeled, time-integrated ARC focal spot of a single beamlet. Dashed contours shown at the entrance and exit apertures of the two cone geometries.  Encircled energy in the cone tip region is 7\% and 20\%  for the 25~\micron{} and 50~\micron{} tip, respectively.}
\label{fig:ExperimentalDiagram}
\end{figure}

The CPC cones were designed to reflect all ARC rays within the maximum angle of 3.8$^\circ$ with the curvature constrained by the output cone tip size. Two cone geometries were fielded with tip diameters of 25 and 50~$\mu$m, lengths of 1.2 and 1.4~mm, and cone opening diameters of 350 and 500~$\mu$m, respectively. The fraction of overall laser energy entering the cones with nominal beam pointing is  84\% and 91\% for the 25~\micron{} and 50~\micron{}  tip, respectively.
The cones, target geometry, and encircled laser spot are schematically shown in Fig.~\ref{fig:ExperimentalDiagram}. The ARC central focal spot is elliptical with a major and minor full width at half maximum of ${\sim}$35 and ${\sim}$15~$\mu$m, respectively and a ${\sim}$25~$\mu$m 1D equivalent. 
%The FWHM spot contains {30\%} of the laser energy, which is larger than the cone tips for each case where for each cone, the tip diameter encircles this region of highest intensity [see Fig.~\ref{fig:LaserSpot}]. 
Cones were attached to a prism-shaped Au target with edge lengths of ${\sim}$2~mm, where the rear facets were normal to electron-positron-proton spectrometers \cite{chen:10E533}. The target dimensions and positioning within the NIF target chamber were identical to that described in Ref.~\cite{Williams2020ElectronsARC} with the exception of the prism height, which was 2~mm for all cone targets.

Six NIF-ARC experiments were performed  to investigate the coupling enhancement between flat and CPC  targets by changing laser energy, pulse duration, and cone geometry. Electrons escaping the target were used as a metric for laser absorption and conversion efficiency by measuring the energy and slope temperature of the electrons with spectrometers located at (90$^\circ$, 78$^\circ$) and (90$^\circ$, 315$^\circ$) in NIF chamber coordinates (58$^\circ$ and 68$^\circ$ from laser axis, respectively). 
Shot parameters and performance metrics for all experiments are given in Table~\ref{table:ExpParams}.
%Four experiments used all four beamlets of ARC, pointed separately onto the flats and pointed together into the cones, and a pulse duration of $\tau_L = 10$~ps and 2.2-2.6~kJ of total energy, except for the low intensity case where 0.6~kJ of energy was used.
The temperature of the electrons was determined by fitting an exponential function to the high-energy portion, from 2-4~MeV to the detection threshold of ${\sim}10^8$~electrons/steradian/MeV. 

\begin{figure}[b]
\begin{center}
\subfigure{\label{fig:FlatVsCone}\includegraphics[width=0\textwidth]{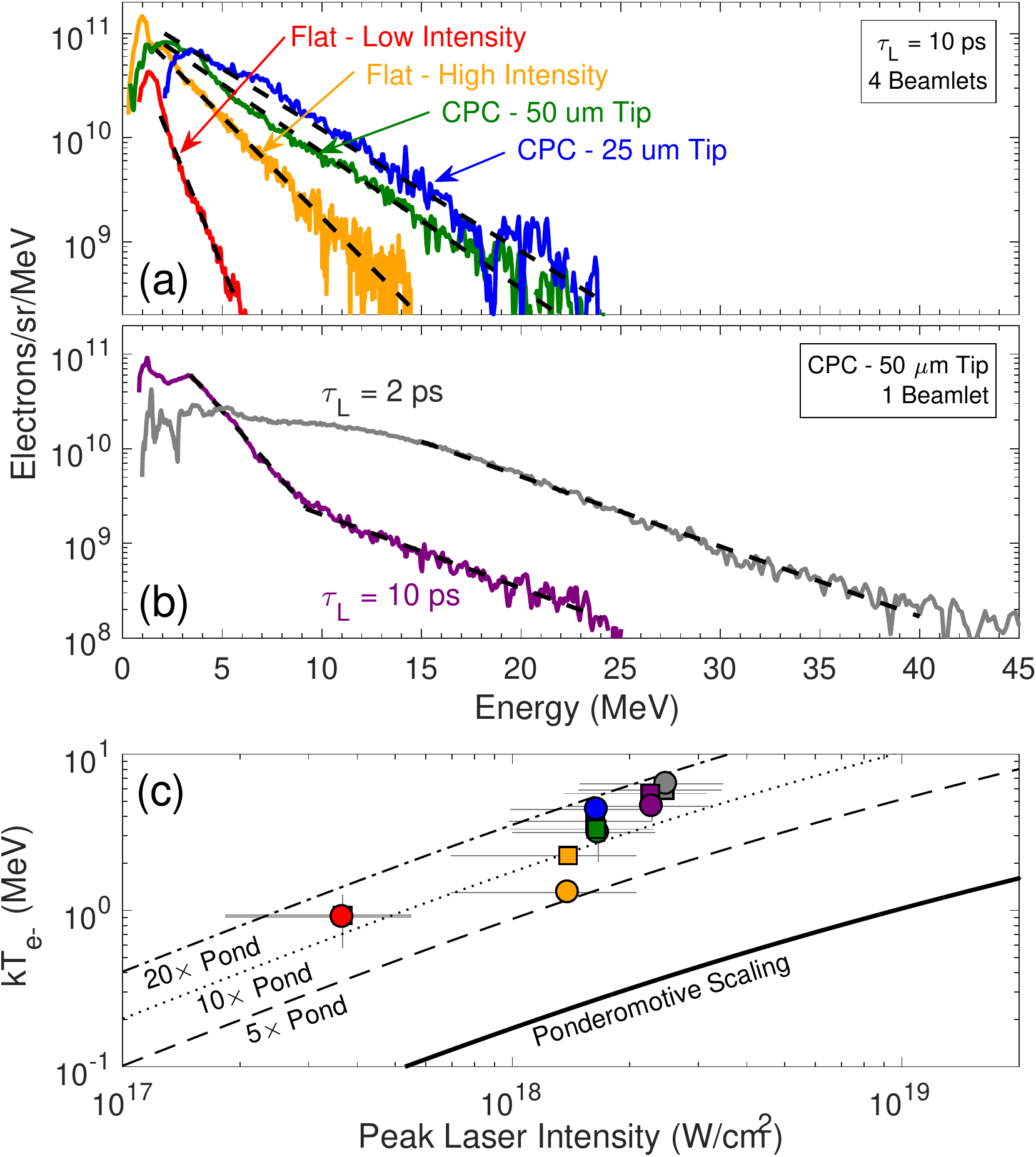}}   \\
\subfigure{\label{fig:10psVs1ps}\includegraphics[width=0\textwidth]{ARC_Cones_AllElectrons_R1_v1-eps-converted-to.pdf}}  
\subfigure{\label{fig:ElectronScaling}\includegraphics[width=0.46\textwidth]{ARC_Cones_AllElectrons_R1_v1-eps-converted-to.pdf}}
\caption{
Electron spectra measured from the (90$^\circ$, 78$^\circ$) line of sight from (a) flat and cone targets using four ARC beamlets at 10~ps pulse duration and (b) 50~$\mu$m cone tip and a single ARC beamlet at 2 and 10~ps pulse duration.
(c) Electron temperatures and ponderomotive scaling (lines) as a function of laser intensity for the (90$^\circ$, 78$^\circ$) and (90$^\circ$, 315$^\circ$) lines of sight designated by square and circle markers, respectively, with colors corresponding to (a) and (b). All experimental parameter and results are given in Table~\ref{table:ExpParams}.}
\label{fig:NEPPS_ElectronSpectra}
\end{center}
\end{figure}

Four experiments used all four ARC beamlets and 10~ps pulse durations to demonstrate scaling and optimize escaping electron performance. 
Results from the flat targets irradiated at a ``low intensity''  and ``high intensity,'' previously reported in Ref.~\cite{Williams2020ElectronsARC}, are compared in Fig.~\ref{fig:FlatVsCone} to the two CPC geometries. %
The electron temperatures measured using cone targets [see Fig.~\ref{fig:ElectronScaling}] were between 12-17 times greater than predicted from ponderomotive scalings, where $kT_{\rm pond} = mc^2 [({1+ I_L \lambda^2 / 1.37 \times 10^{18}})^{1/2}- 1]$. 
%The electron temperatures from each line of sight are shown in Fig.~\ref{fig:ElectronScaling} with multipliers on the ponderomotive scaling.
The 25~$\mu$m-tip CPC target increased the observed electron temperature by a maximum of 3.4$\times$ and coupling efficiency by 7.5$\times$ compared to flat targets for nominally identical input laser conditions. Higher temperatures and conversion efficiencies were observed using a 25~$\mu$m-tip CPC compared to a 50~$\mu$m-tip. A larger electron flux was observed in the (90$^\circ$, 315$^\circ$) line of sight for most experiments, which is likely due to a dynamic target charging effect originating from the asymmetric prism geometry.

\begin{table*}[t]
\begin{center}
\begin{ruledtabular}
\begin{tabular}{|cccccc|cc|cc|} 
  \multicolumn{6}{|c|}{Shot Parameter} 															&\multicolumn{2}{c|}{C.E. (mJ/sr/kJ) } 			& \multicolumn{2}{c|}{kT (MeV)}  						\\
NIF Shot ID 	& Description 		 	& BL		& $E_L$ (J) 	& $\tau_L$ (ps)		& $I_{18}^{\rm peak}$ 	& (90$^\circ$, 78$^\circ$)			& (90$^\circ$, 315$^\circ$) 	& (90$^\circ$, 78$^\circ$) 		& (90$^\circ$, 315$^\circ$) 	\\ \hline
N170514-002 	& Flat - Low Intensity 	& 4		& \,\,\,649		 & 10 			& 0.4 				& \,\,\,$3.9  \pm 0.8$ 				& $11 \pm 1$				& $0.93 \pm 0.34$			& $0.91 \pm 0.34$ 			\\
N170517-003	& Flat - High Intensity 	& 4		& 2371		 & 10 			& 1.4 				& $30 \pm 6$ 					& $35 \pm 7$ 				& $2.2 \pm 0.2$ 			& $1.3 \pm 0.1$ 			\\
N180515-001 	& CPC - 50 $\mu$m Tip 	& 4		& 2172		 & 10 			& 1.7 				& $100 \pm 20$ 				& $130 \pm 30$			& $3.3 \pm 0.4$ 			& $3.1 \pm 1.1$ 			\\
N181204-001 	& CPC - 25 $\mu$m Tip	& 4		& 2600		 & 10 			& 1.6 				& $130 \pm 30 $ 				& $240 \pm 50$ 			& $3.7 \pm 0.5$ 			& $4.4 \pm 0.8$ 			\\
N180514-004  	& CPC - 50 $\mu$m Tip 	& 1		& \,\,\,545		 & 10 			& 2.3					& $220 \pm 40$ 				& $150 \pm 30$ 			& $5.6 \pm 2.1$ 			& $4.6 \pm 1.8$ 			\\
N181204-002 	& CPC - 50 $\mu$m Tip	& 1		& \,\,\,272		 & \,\,\,2 			& 2.5 				& $2000 \pm 390$ 				& $1600 \pm 320$ 			& $5.9 \pm 1.3$ 			& $6.5 \pm 2.1$ 			\\
\end{tabular}
\end{ruledtabular}
\end{center}
\caption{Shot parameters and performance metrics for all experiments. Energy ($E_L$) is the total energy of all beamlets (BL). Average peak intensity for all beamlets, $I_{18}^{\rm peak}$, is reported in units of $10^{18}$~W/cm$^2$. Conversion efficiency (C.E.) defined as the total electron beam energy between 3~MeV and detection threshold at a line of sight, normalized to the laser energy in kJ.}
\label{table:ExpParams}
\end{table*}%

%3.67E+17
%1.39E+18
%1.65E+18
%1.63E+18
%2.46E+18
%2.26E+18

%The (90$^\circ$, 78$^\circ$) line of sight showed a temperature increased from $2.2 \pm 0.2$~MeV to $3.3 \pm 0.5$~MeV and $3.5 \pm 0.5$~MeV and the energy in the electrons from $0.10 \pm 0.02$~J/sr to $0.23 \pm 0.05$~J/sr and $0.36 \pm 0.07$~J/sr for the flat, 50~$\mu$m-tip, and 25~$\mu$m-tip CPC, respectively. The (90$^\circ$, 315$^\circ$) line of sight showed a temperature increased from $1.3 \pm 0.1$~MeV to $3.1 \pm 1.1$~MeV and $3.7 \pm 0.5$~MeV and the energy in the electrons from $0.18 \pm 0.04$~J/sr to $0.26 \pm 0.05$~J/sr and $0.64 \pm 0.13$~J/sr for the flat, 50~$\mu$m-tip, and 25~$\mu$m-tip CPC, respectively. 
%
%

Two experiments used a 50~$\mu$m-tip CPC cone with a single beamlet of ARC at either the transform limited pulse duration ($\tau_L = 2$~ps) or $\tau_L = 10$~ps to explored the pulse-length dependent absorption behavior of the cones [see Fig.~\ref{fig:10psVs1ps}]. The 10~ps data shows a clear two-temperature spectrum with low- and high-energy temperatures, the later of which is comparable to the $\tau_L = 2$~ps electron temperature. Normalizing to the incident laser energy, the 2~ps experiment had ${\sim}10{\times}$ more energy in the observed electrons than the 10~ps experiment and suggests that the later parts of pulse are less efficient or that the plasma evolution degrades the performance of the CPC target.
%
%longer pulse duration degrades the performance of the CPC target.
%
Interestingly, the average number of observed electrons in the 10~ps, four-beamlet experiment is only twice that of the single-beamlet case. Beamlet pointing jitter ($\pm$30-40~$\mu$m) and a systematic timing error at the time of the experiments (${\sim}15$~ps) are hypothesized to contribute to  this degradation, where beamlet co-timing is likely the largest effect as the late pulses would be interacting with a modification to the initial conditions of the cone tip and walls.
The CPC geometry is designed to loosen pointing tolerance constraints whereas correcting the timing error may lead to significantly higher conversion efficiencies and electron temperatures than presented here. 
%We conclude that the four-beamlet, $\tau_L = 2$~ps, and 25~$\mu$m-tip CPC configuration would optimize the conversion efficiency from laser to electron and secondary particles.

\begin{figure}[b]
\includegraphics[width=0.46\textwidth]{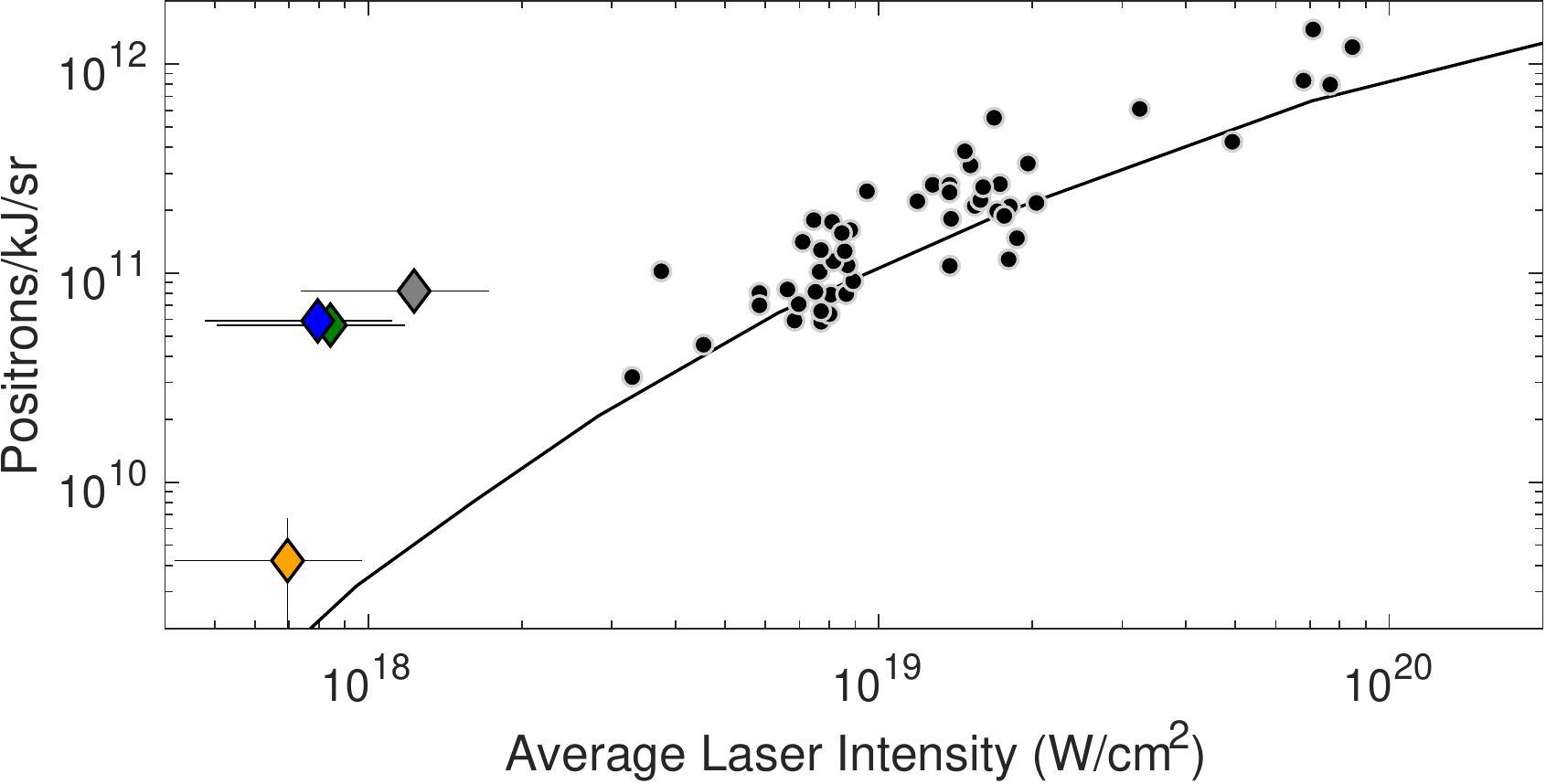}
\caption{Measured positron yield as a function of average laser intensity over the spot size (1/$e^2$). Diamond markers indicate current experiments. Scaling model (line) and data from other facilities (dots) from Ref.~\cite{PhysRevE.79.066409} and Ref.~\cite{PhysRevLett.114.215001}, respectively.
}
\label{fig:PositronScaling}
\end{figure}

Positrons were observed on four of the experiments and are compared to data collected at other laser facilities and an analytic model in Fig.~\ref{fig:PositronScaling}. The positron yield from the flat target shot agrees well with the model while the CPC targets show a production efficiency nearly an order of magnitude greater than established scaling. This result is consistent with the observed increase in the number of electrons with energies ${>}$10~MeV, which are the dominant driver of pair production \cite{nakashima2002numerical}.
 Notably, these pair yields are normally associated with lasers with average intensities of ${\sim}10^{19}$~W/cm$^2$.

%To quantify the performance and investigate the laser interaction inside the cone

%\fix{Fit in somewhere:} We argue that this agreement is evidence of geometric focusing by the cone at early times before the cone wall density blows down and causes incident rays to refract, moving the focus away from the cone tip. Parallel to this effect is the plasma confinement and modification of the scale length that the cone provides, which can lead to the efficient production of superponderomotive electrons \cite{PhysRevLett.109.195005}. The interdependency and relative contribution of these two processes for geometry and laser conditions will be the subject of followup investigations and advanced simulations. 

%\fix{2D PIC simulations:}

CPC optics used in low-power applications will geometrically focus and intensify the light at the cone tip. With high-power lasers, however, absorption on the wall, and the resulting change in reflectivity and geometry due to the presence of plasma, will modify the behavior of light within the cone. Previous work broadly hypothesized that CPC targets enhanced coupling through a combination of geometric focusing and more efficient absorption at the cone tip due to confinement of underdense plasma \cite{MacPhee2020CPC}.

Here, we quantify this hypothesis and elucidate an understanding of the mechanisms responsible for the enhanced performance between flat and CPC cone targets by using two dimensional hydrodynamic and PIC simulations. % 
Simulations of the 10~ps experiments were computationally prohibitive and we instead focus only on the absorption mechanisms using shorter pulse durations.
The laser-plasma interactions were modeled using the hybrid PIC code Chicago \cite{Thoma2017Hybrid}. The laser pulse was initialized with a 1~ps FWHM sin$^2$ temporal profile and two-component radial approximation to the experimental spot with a 26 and 120~$\mu$m FWHM containing 33\% and 67\% of the incident energy, respectively, resulting in a vacuum peak intensity of $2.25 \times 10^{18}$~W/cm$^2$.
The simulation box had 10 cells per wavelength and 15 steps per optical cycle and was $\pm$150~$\mu$m in the transverse dimension and $\pm$300~$\mu$m in the axial direction, centered at the initialized target surface.

\begin{figure}[t]
\subfigure{\label{fig:Sim_Density}\includegraphics[width=0\textwidth]{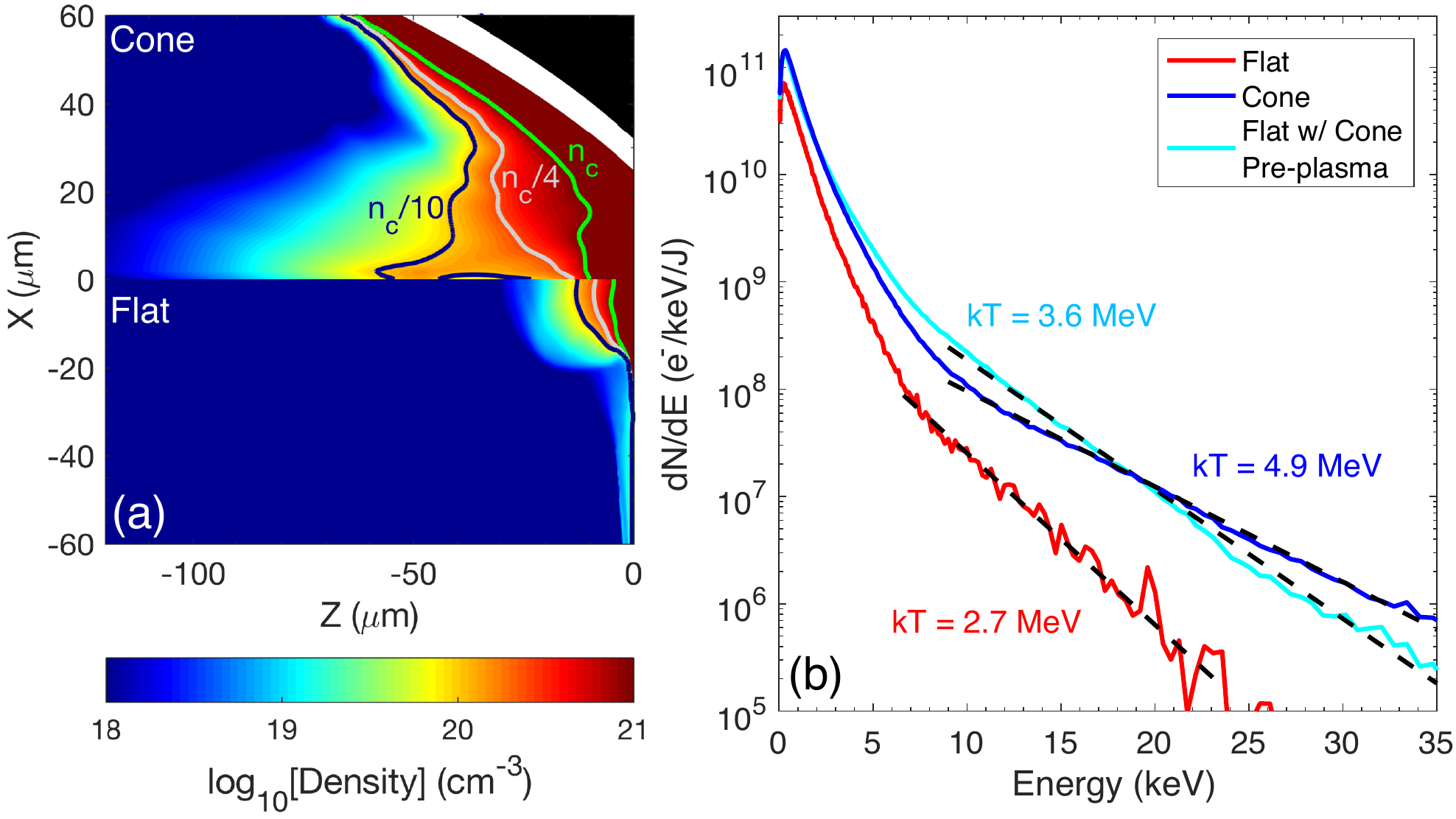}}  
\subfigure{\label{fig:Sim_ElectronTemp}\includegraphics[width=0.48\textwidth]{Sim_Density_ElectronTemp_v3.pdf}}  
\caption{
(a) Density maps from hydrodynamic simulations for cone (upper) and flat (lower) targets with contours corresponding to $n_c$, $n_c/4$, and $n_c/10$. White region shows the initial location of the cone wall.
(b) Electron spectra from cone, flat, and flat with cone preplasma. Exponential temperatures are fit to the high energy region of the spectra.
}
\label{fig:SimDensity_ElectronTemp}
\end{figure}

The preformed plasma evolution was  calculated by the radiation hydrodynamic code HYDRA  \cite{doi:10.1063/1.1356740} using the measured ARC focal spot distribution and laser prepulse temporal profile and assumed 1\% absorption (99\% reflectivity) for laser rays interacting with overdense material.  The hydrodynamic simulations  [Fig.~\ref{fig:Sim_Density}] show  enhanced plasma growth inside the cone tip.  The volume of plasma density between $0.1 < n_e/n_c < 1$, where the laser is heavily absorbed, increases by 10${\times}$ from the flat to the cone simulation. 
The additional plasma is due to laser rays in the cone tip region having several chances to deposit energy whereas those rays reflecting from the flat exit the system after typically one bounce. 
The initial density scale length seen by the main laser pulse along $z$ increases from 4~$\mu$m in the flat case to ${\sim}$15~$\mu$m in the cone case. The cone walls confine the plasma where the expansion is not limited to only the transverse spot size of the laser.
The simulated time-integrated electron spectra are shown in Fig.~\ref{fig:Sim_ElectronTemp} and reproduce the trends observed in the data where the cone target has a higher conversion efficiency into energetic electrons (${>}5$~MeV) compared to the flat target by a factor of four. The electron temperature in the flat case is 2.7~MeV and nearly doubles to 4.9~MeV for the cone target.

\begin{figure}[b]
\subfigure{\label{fig:Sim_IntensityNoCone}\includegraphics[width=0\textwidth]{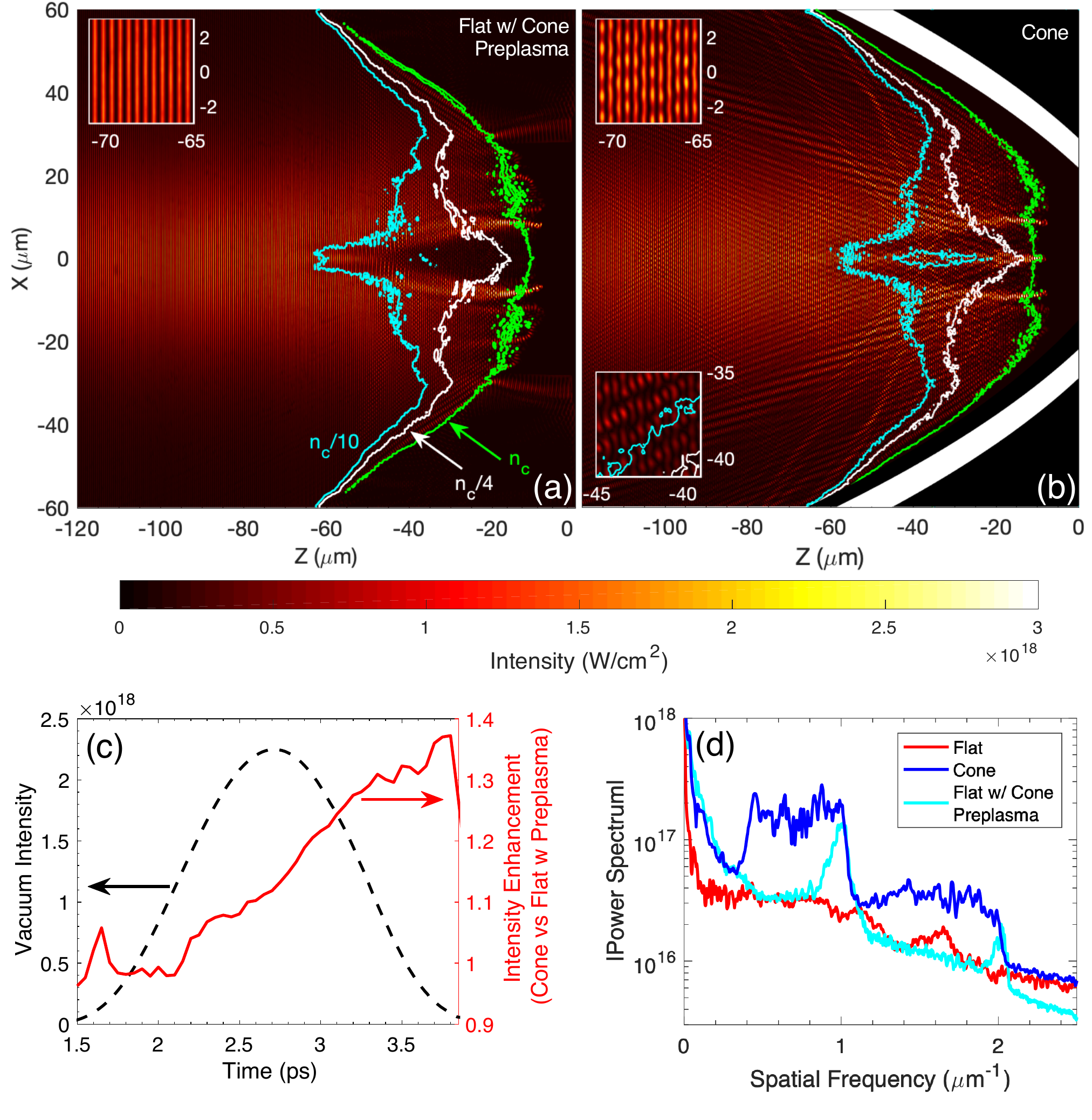}}  
\subfigure{\label{fig:Sim_IntensityCone}\includegraphics[width=0\textwidth]{SimulationPlot_Intensity_v3.pdf}}  
\subfigure{\label{fig:Sim_AvgIntensityTip}\includegraphics[width=0\textwidth]{SimulationPlot_Intensity_v3.pdf}}  
\subfigure{\label{fig:Sim_FourierTransform}\includegraphics[width=0.48\textwidth]{SimulationPlot_Intensity_v3.pdf}}   
\caption{
Field intensity plots for (a) flat with cone preplasma and (b) cone targets with contours corresponding to $n_c$, $n_c/4$, and $n_c/10$. (c) Input laser intensity (dashed) and ratio of intensities for test and cone case averaged between $-30 < x (\mu {\rm m}) < 30$ and within densities between $0.1 < n_e/n_c < 0.25$ (solid).
(d) Fourier transform of transverse Poynting flux summed over the axial dimension for all cases.
}
\label{fig:SimIntensity_ConeNoCone}
\end{figure}

%temperature increases from 2.7~MeV to 4.9~MeV for the flat and cone, respectively. 
%In the case of the cone, the plasma profile is a contribution of material moving on axis as well as the central spot on axis.

%
% large differences between the plasma that develops in the case of a flat and cone geometry due to the laser prepulse [see Fig.~\ref{fig:Sim_Density}] where at very early times, light is efficiently focused to the cone tip and increases the local intensity at the cone tip.  

%
% Mark: In the case of the cone, the plasma profile is a contribution of material moving on axis as well as the central spot on axis.
%Each case shows a two dimensional structure; due to the laser central spot in the flat case, while in the cone case plasma expansion from the nearby cone walls is a large contribution .  

% Mark: the simulated time integrated spectra
%The simulated time-integrated electron spectra are shown in Fig.~\ref{fig:Sim_ElectronTemp} and reproduce the trends observed in the data where conversion efficiency into energetic electrons more than doubles and the temperature increases from 2.7~MeV to 4.9~MeV for the flat and cone, respectively. 
% Mark: temperature in the flat is 2.7 and increases to 4.9 for the cone

%The increase of these metrics can be accomplished through laser intensification or more efficient absorption mechanisms in the larger underdense plasma volume.

% performance metrics 
% Mark: TO see if the enhancement in temperature and CE is due to the added plasma, ...

To determine whether the enhancement in temperature and conversion efficiency is due to the added plasma alone, a simulation of the flat target was performed with the initial preplasma from the cone scenario artificially inserted and plasma with density ${>}0.9 n_c$ outside of the cone tip region removed.
This enables us to look at the effect of the cone's preplasma alone, without the presence of cone walls that reflect light.
The cycle-averaged Poynting flux for this test scenario and the cone are shown  near peak intensity in Fig.~\ref{fig:Sim_IntensityNoCone} and \ref{fig:Sim_IntensityCone}, respectively.
In the test case, the laser maintains a near-plane wave profile prior to reaching densities of ${\sim}0.1 n_c$, after which typical modulations are observed caused by Raman backscatter and filamentation, where  the laser self-focuses to intensities ${\sim}$4 times the vacuum intensity \cite{doi:10.1063/1.4729333}. 
This is in contrast to the cone case where intensity and phase modulations are visible at very large distances from the tip (see figure insets). 
%
%due to cone-reflected waves interfering with the incoming light.  
%
Near the cone wall and throughout the cone tip where near-critical density exists, the large angle between incident and reflected light produces localized islands of high intensity with a turbulent wavenumber-spectrum. Electrons that traverse such islands can dephase from the laser and will experience a short period of acceleration or deceleration. This process repeats itself as electrons cross a large volume of these intensity islands and, on average, significant acceleration can occur above ponderomotive expectations.

The effect of plasma in the test case is primarily to increase conversion efficiency to a level comparable to the cone simulation, yet it only accounts for half the temperature increase [see Fig.~\ref{fig:Sim_ElectronTemp}]. Therefore, the reflectivity of the cone wall has only a marginal affect on the yield of energetic electrons. This result suggests that the increase in conversion efficiency experimentally observed when moving from the 50~$\mu$m tip to the 25~$\mu$m tip was due to the change in plasma conditions and not optical focusing. 
The cone wall reflections are shown to increase the average intensity at the cone tip by 5-30\% during the peak of the laser pulse [Fig.~\ref{fig:Sim_AvgIntensityTip}], which will only marginally influence the electron temperature assuming a square root dependence on laser intensity (ponderomotive).
The lack of strong geometric convergence due to the presence of cone walls in 2D suggests a 3D simulation would be unlikely to explain the increase in intensity required to account for the large temperature increase. 
The presence of plasma at the cone tip disrupts the ability of the cone to concentrate intensity as it might in a low-power application.

We conclude that the enhanced acceleration is caused by the additional laser field non-uniformities that cover a significantly larger volume for the cone case than either the flat or test cases. These turbulent fields create collections of isolated wavelets, rather than uniform plane waves, with high spatial frequencies. 
The presence of transverse turbulence in the laser field can be quantified by observing the Fourier transform of the Poynting flux in the transverse dimension at peak intensity [see Fig.~\ref{fig:Sim_FourierTransform}]. The presence of plasma creates harmonics of the central wavelength in the transverse dimension yet a large bandwidth of spatial frequencies is only observed when light is reflecting from the cone walls. 
This increase in bandwidth allows electrons to more easily dephase and therefore undergo non-adiabatic acceleration more readily \cite{sherlock2020electron}.
\fix{Ideal conditions were considered for the simulations which included only axial propagation. For scenarios with more than one beamlet and realistic pointing jitter, more laser energy will hit the cone wall at grazing incident and refract at a higher angle.  While this refraction reduces the focusing strength of the optic, it could mitigate against performance degradation due to pointing misalignment.}

In summary, we demonstrate that CPC cones can be used as robust target components to increase the conversion efficiency into MeV electrons by nearly an order of magnitude and electron temperature by ${>}3{\times}$ compared to flat targets.
Hydrodynamic and PIC simulations reveal that the increases in conversion efficiency are primarily due to additional preformed plasma at the cone tip, whereas the temperature increase is due to a combination of the longer scale length plasma and enhanced acceleration mechanism driven by laser turbulence that develops from reflections from the cone walls. 
At these intensities, CPC cones do not act as geometric optics that would intensify light at the cone tip. 
The platform developed here can be used to optimize many short pulse applications requiring enhancements to relativistic electron acceleration such as secondary ion or photon particle sources, the generation of high energy density environments, and pair production. 
The ability of the cone to generate long plasma scale lengths, overcoming the limitations of finite-sized focal spots, as well as harnessing the energy contained in the wings of the beam to initiate turbulent laser fields suggests these CPC targets could be used with a wide range of laser facilities.

%\section{Acknowledgements}
We thank the ARC Laser team and NIF Experimental Operations team for implementing these experiments. 
We gratefully acknowledge the NIF Discovery Science Program for the experimental allocation.
This work was performed under the auspices of the U.S. Department of Energy by the Lawrence Livermore National Laboratory under Contract DE-AC52-07NA27344 and funded by the LLNL LDRD program under tracking code 17-ERD-010. Target fabrication was performed with General Atomics under contract DE-NA0001808.

\end{document}